\def \SAIT #1 #2 {{\em Mem.\ Soc.\ Astron.\ It.\/} {\bf #1}, #2}
\def \MESS #1 #2 {{\em The Messenger\/} {\bf #1}, #2}
\def \ASTRNACH #1 #2 {{\em Astron. Nach.\/} {\bf #1}, #2}
\def \AAP #1 #2 {{\em Astron. Astrophys.\/} {\bf #1}, #2}
\def \AAL #1 #2 {{\em Astron. Astrophys. Lett.\/} {\bf #1}, L#2}
\def \AAR #1 #2 {{\em Astron. Astrophys. Rev.\/} {\bf #1}, #2}
\def \AAS #1 #2 {{\em Astron. Astrophys. Suppl. Ser.\/} {\bf #1}, #2}
\def \AJ #1 #2 {{\em Astron. J.\/} {\bf #1}, #2}
\def \ANNREV #1 #2 {{\em Ann. Rev. Astron. Astrophys.\/} {\bf #1}, #2}
\def \APJ #1 #2 {{\em Astrophys. J.\/} {\bf #1}, #2}
\def \APJL #1 #2 {{\em Astrophys.. J. Lett.\/} {\bf #1}, L#2}
\def \APJS #1 #2 {{\em Astrophys. J. Suppl.\/} {\bf #1}, #2}
\def \APSS #1 #2 {{\em Astrophys. Space Sci.\/} {\bf #1}, #2}
\def \ASR #1 #2 {{\em Adv. Space Res.\/} {\bf #1}, #2}
\def \BAIC #1 #2 {{\em Bull. Astron. Inst. Czechosl.\/} {\bf #1}, #2}
\def \JSQRT #1 #2 {{\em J. Quant. Spectrosc. Radiat. Transfer\/} {\bf #1},#2}
\def \MN #1 #2 {{\em Mon. Not. R. Astr. Soc.\/} {\bf #1}, #2}
\def \MEM #1 #2 {{\em Mem. R. Astr. Soc.\/} {\bf #1}, #2}
\def \PLR #1 #2 {{\em Phys. Lett. Rev.\/} {\bf #1}, #2}
\def \PASJ #1 #2 {{\em Publ. Astron. Soc. Japan\/} {\bf #1}, #2}
\def \PASP #1 #2 {{\em Publ. Astr. Soc. Pacific\/} {\bf #1}, #2}
\def \NAT #1 #2 {{\em Nature\/} {\bf #1}, #2}

\newcommand{\be}{\begin{equation}}
\newcommand{\ee}{\end{equation}}
\newcommand{\etal}{{\em et al.}}

\def\picture #1 by #2 (#3){
  \vbox to #2{
    \hrule width #1 height 0pt depth 0pt
    \vfill
    \special{picture #3} 
    }
  }
\def\scaledpicture #1 by #2 (#3 scaled #4){{
  \dimen0=#1 \dimen1=#2
  \divide\dimen0 by 1000 \multiply\dimen0 by #4
  \divide\dimen1 by 1000 \multiply\dimen1 by #4
  \picture \dimen0 by \dimen1 (#3 scaled #4)}
  }
\def\boxit#1{$$\vbox{\hrule\hbox{\vrule\kern3pt
     \vbox{\kern3pt \hbox to 13cm{\hfil }\vskip
     #1cm\kern3pt} \kern3pt \vrule}
     \hrule}$$}

\documentclass{article}
\usepackage{epsfig,graphicx}
\begin{document}
\input epsf.sty
\title{   Tau air showers detectability with GLAST}
\author{ Daniele Fargion$^1$ $^2$ \\
{\em $^1$Dept. of Physics,  Univ. of Roma ''Sapienza''  and INFN Roma, Italy} \\
{\em $^2$Dept. of Physics,  Univ. of Roma ''Tor Vergata''  and INFN Roma 2, Italy}}
\maketitle
\baselineskip=11.6pt
\begin{abstract}

We show that with GLAST there will be the possibility to detect, within the UHECR skimming the Earth atmosphere,
 the showers generated by very high energy upward and horizontal Tau. The effective area, thanks to the large area covered by the showers at 550 Km, is  less than that of AUGER, but its efficiency is comparable because the lower detection threshold and the consequent event rate may lead to a few  EeV and-or few Glashow resonant signals within a decade.
\end{abstract}

\baselineskip=14pt

\section{Introduction}

Ultra high energy neutrinos UHE  $ \nu_{\tau}$, $ \bar\nu_{\tau} $
and $ \bar{\nu}_e $ at EeV's up to GZK energies ($\ge 10^{19}$ eV)
can hit the earth crust at the horizon leading to UHE $\tau$ which may decay in flight at high altitude.
The consequent UHE  air showers  might be observable by next generation gamma-ray space missions like GLAST.

Here we show the expected  fluence and time signature considering two different complementary signals:
the upward $\tau$ air shower (UpTau) near the vertical at PeV energies and the horizontal $\tau$ air shower ( HorTau )
at 1.4 10$^{19}$ eV.

\subsection{Upward $\tau$ air shower}

Assuming a given altitude ($h_1 \sim 575 Km$) for the circular
orbit of the satellite, the distance between the detector and the
edge of the earth crust can be written as
\begin{equation}
{d_{1U}=(R+h_{1})\sin(\theta_{1U})- {\sqrt{(R+{h_{1}})^2 \cdot {\sin}^{2}(\theta_{1U})-[(R_\oplus  + h_1)^2 - R_\oplus^2]}}}
\end{equation}
where $\theta_{1U}\sim 70^o$ is the angle of the shower from the horizontal. In first approximation
$${d_{1U}  \sim h_1 / \sin \theta_{1U}  \sim 612 Km }$$

A pictorial view of the detection method is shown in figure~\ref{glastscheme1}.

\begin{figure}
\vspace{0cm}
    \epsfig{file=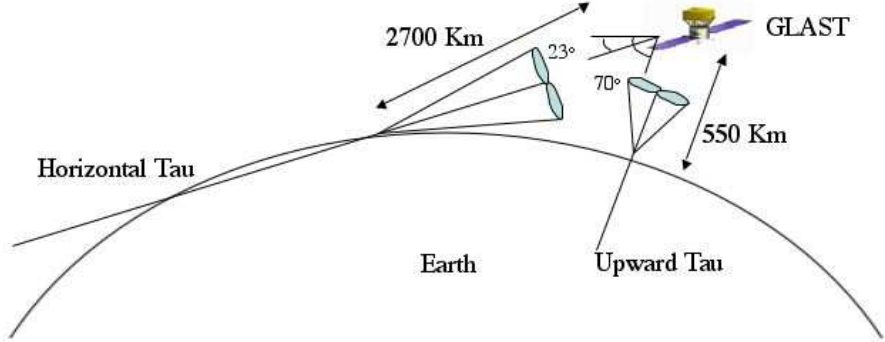,width=1\textwidth,angle=0,clip=}
\caption{\label{glastscheme1} \it A schematic picture (not in scale) of the possible detection method }\label{glastscheme1}
\end{figure}

Now we can calculate the area $ A_U $ of the corresponding  front
of the  upward $\tau$ showers that is given by :

$$ A_U = {\pi \over 4} \Delta \theta_{sh}^2 d_{1U}^2 \simeq  90 Km^2 $$

where $\Delta \theta_{sh} \sim 1^o$ is the typical opening angle
for showers.

The lateral density profile is, of course, more dense near the inner part, and here we assume that $\sim 90 \%$
of the gammas are contained in a narrow angle of $1/4 $ of degree, leading to a reduces  area:

 $$A_{Ur}=5.62  Km^2$$

This areas allow us to calculate  the secondary gamma-ray flux.

We considered     $\nu_\tau$'s with a  primary energy of the order of  $\sim 4  \cdot  10^{15}$ eV  because at greater energies they are suppressed by the earth opacity, while at lower energies the cross section and the $\tau$ propagation length are smaller \cite{tau}.
Indeed the probability $P(\theta,\, E_{\nu})$ of escaping from the earth is approximately
\begin{equation}
P(\theta_{1U},\, E_{\nu}) \simeq e^{\frac{-2R_\oplus \sin
\theta_{1U}}{R_{\nu_{\tau}}(E_{\nu})}} (1 - e^{-
\frac{R_{\tau}(E_{\tau})}{R_{\nu_{\tau}}(E_{\nu})}}) \, .
\end{equation}

where $\theta_{1U}$ is nearly the complementary angle of the direction of the upcoming $\tau$ angle with the zenith, $R_{\tau}$ is the interaction length of the $\tau$
and $R_{\nu_{\tau}}$ is the $\nu$ interaction length.

\begin{figure}[htbp]
  \epsfig{file=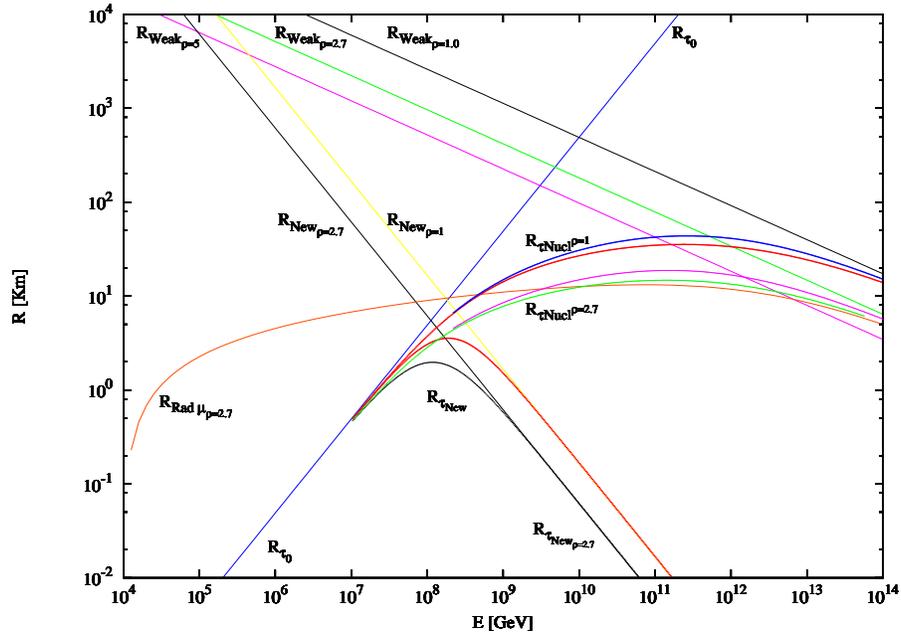 ,width=1\textwidth,angle=0,clip=}
 \caption{Lepton $\tau$ (and $\mu$)
Interaction Lengths for different matter densities: $R_{\tau_{o}}
=  c\cdot {\tau_{\tau}}\cdot {\gamma_{\tau} } $ is the free $\tau$
length,$R_{\tau_{New}}$ is the New Physics TeV Gravity
interaction range at corresponding densities,
$R_{\tau_{Nucl}\cdot{\rho}}$ ,\cite{tau}, see also
\cite{Dutta et al.2001}, is the combined $\tau$ Ranges keeping
care of all known interactions and lifetime and mainly the
photo-nuclear interaction. There are two slightly different split
curves (for each density) by two comparable approximations in the
interaction laws. Note also the neutrino interaction lenghts
above lines $R_{Weak{\rho}}= L_{\nu}$ due to the electro-weak
interactions at corresponding densities (see also \cite{Gandhi et
al 1998})  \cite{tau}.} \label{fig4}
\end{figure}

The $\tau$ energy is typically  $\sim 20 \% $ less then the $\nu_\tau$ one's.

The number of gamma $ N_{\gamma s}$ with energies around 100 MeV in the showers is in first approximation
$$ N_{\gamma s} \sim { E_\tau  \over E_c} \sim  4 \cdot 10^6   $$

for $ E_c =100MeV$ with the assumption of the energy
equi-partition between $\gamma$, electron pairs ($\sim$ 33\% each
component), as well as taking in account a partial ($\sim$ 33\%)
opacity of the atmosphere for the $\tau$ shower.

The number of photons per unit reduced area at the altitude of
GLAST is then :

$$\Phi_{\gamma r} = {N_{\gamma s} \over A_{Ur}} =  7.14 \cdot 10^{-5}
cm^{-2}$$ in the inner$\Delta \theta = {1/4} ^o$ core and
$$\Phi_\gamma =  4.4 \cdot 10^{-6} cm^{-2}$$ in the wider
$\Delta \theta = 1^o$ cone shower.

The characteristic time structure of the shower is $ t_s \sim   L_s /  c  \ge 10^{-4} ~s   $

where $L_s $ is the shower attenuation length at altitude  $\sim
23 Km$, where upward $\tau$ take place. (see ref.\cite{tau2}).

Assuming the lateral GLAST detector area, $A=  1.3 \cdot 10^4
cm^2$, an efficiency  $\eta = 0.5 $, the total effective area  is
$ A_{eff} = A \cdot \eta \cdot cos\theta  = 2.3 \cdot 10^3 cm^2$
the number of photons for each event is, respectively for narrow
and large view angle :

$$  N_{\gamma r} (E \sim 100 MeV) =  \Phi_\gamma \cdot A_{eff} \sim 0.16  $$
$$  N_\gamma (E \sim 100 MeV)   \sim 10^{-2}  $$

So we conclude that GLAST can measure upward $\tau$ only in
coincidence with the GRB monitor approximately one over 6 upward
$\tau$ showers. Therefore the high energy $\gamma$ detection alone
is not an effective way to discriminate upward Tau Air-Showers
(UpTaus) by GLAST.

%




\subsection{Horizontal $\tau$ air shower}

We can now use the above procedure to calculate the rate of events for the Horizontal $\tau$ air shower.
The distance between the detector and the edge of the earth crust is in this case

 $d_{hH}=( 2R_\oplus h_1)^{1/2} \cdot  (1+ {h_1 \over 2 R_\oplus})^{1/2} \sim 2768$ Km
 for the same altitude of 575 Km and where  the angle of the shower from the horizontal is
  $\theta_{hH} = \arctan(( {2h_1 \over R_\oplus} )^{1/2} \cdot  (1+ {h_1 \over 2 R_\oplus})^{1/2} )\sim 23.5^o$.
  However the Tau decay in flight and the HorTau appearence takes place at great distance ($ \simeq 600$ km) from
  the Earth and the HorTau Shower has a characteristic distance of
  ($\simeq 200$ km) making the real distance from the Shower front
  to the satellite reduced to  $d_{hHorTau} \sim 2500$ km.

Now we can calculate the area of the corresponding  front of the showers given by :
$$ A_H = {\pi \over 4} \Delta \theta_{sh}^2 d_h^2 \sim  1510  Km^2
$$
This area is comparable with future AUGER experiment area.
In analogy to previous UpTaus scenario we also consider inner
Shower cone of a nominal beam angle $1/4$ of degree obtaining a
reduced area $$ A_{H r} = {\pi \over 4} \Delta \theta_{sh}^2 d_h^2
\sim 94.4  km^2$$.

This areas allow us to calculate  the secondary gamma-ray flux.

The optimal observable primary neutrino energy is  1.1 $ 10 ^{19}$
eV because of the earth crust slant  depth  combined with the
horizontal atmospheric opacity \cite{tau}.

The number of gamma $ N_\gamma$ with energies around 100 MeV in the showers is in first approximation
$$ N_\gamma \sim { E_\tau  \over E_c} \sim  3.3  \cdot 10^{10} $$
with the same assumption of the energy equi-partition between
$\gamma$, electron pairs but without the  opacity of the
atmosphere for the $\tau$ shower because in this case we are at
the maximum of the shower with nearly no atmospheric suppression.

The number of photons per unit area (or reduced area) at the
altitude of GLAST is then :

$$\Phi_\gamma = {N_{\gamma s} \over A_{Ur}} =  2.18 \cdot 10^{-3}  cm^{-2}$$
$$\Phi_{\gamma r} = {N_{\gamma s} \over A_{Ur}} = 3.5 \cdot 10^{-2}  cm^{-2}$$

The characteristic time structure of the shower is $ t_s \sim
L_s /  c  \ge 10^{-3} ~s   $

where $L_s \simeq 200 $ km,  is the shower attenuation length at
high altitude $\sim 23 $km, where air is much diluted (see
ref.\cite{tau2}).

Assuming  as before  the lateral area of the detector $A= 1.3
\cdot 10^4 cm^2$, an efficiency  $\eta = 0.5 $, the total
effective area  is $ A_{eff} = A \cdot \eta \cdot \cos\theta = 0.6
\cdot 10^4 cm^2$ the number of photons for each event is for
$\Delta \theta_{Sh}  = 1^o, {1/4}^o$:

$$  N_\gamma (E \sim 100 MeV)  \sim  13.1  $$
$$  N_{\gamma r} (E \sim 100 MeV)  \sim  210  $$

\subsection {HorTau Event rate in GLAST}
The number of events may be estimated by scaling  the EUSO
experiment event rate at the horizons, keeping care of the
different beaming angle and of the different horizontal area and
duty cycle life-time $\eta_{EUSO} \simeq 0.1$ respect the  GLAST
one $\eta_{GLAST} \simeq 1$ , for a nominal three years of
recording. These event rate are scaled assuming a minimal,
guaranteed GZK ( Greisen, Zatsepin, Kuzmin) neutrino fluence
$\Phi_{\nu GZK} \simeq \Phi_{UHECR} \simeq  3 \cdot 10 ^{-18} cm
^{-2} s^{-1} sr^{-1}$ produced by observed Ultra High Cosmic
Rays, UHECR, during their photopion scattering on Cosmic Big Bang
Radiation within the GZK cut-off volumes:
$$ {N_{GLAST}} = \frac{A_{GLAST}}{A_{EUSO}}\frac{1}{360^o} \cdot
\frac{1}{\eta_{EUSO}} {N_{EUSO}}\frac{1}{2} \sim 0.398 {N_{EUSO}
\sim {15}\leftrightarrow{30}}  $$

The consequent reduced area (narrower beamed) event number is:

 $$ {N_{GLAST r}} = \frac{A_{GLAST}}{A_{EUSO}}\frac{1}{5760^o} \cdot
\frac{1}{\eta_{EUSO}} {N_{EUSO}}\frac{1}{2} \sim 0.0248 {N_{EUSO}
\sim 1\leftrightarrow 2} $$

\subsection{HorTaus versus Other High Altitude Showers}
   Among these Upward-Horizontal Showers by $\tau$ we must
   consider the competitive signals of more common and known UHECR showers
   at horizons: Horizontal High Altitude Shower Hias (\cite{tau2} ) are observed by satellites
   above the horizons ($\theta \geq {0.8}^o$) and they behave
   as a  background signal respect to HorTau below the Horizons ($\theta \leq {0.05}^o$).
   Indeed their event number in three years  (at the same GZK energies $10^{19}$ eV, and flux $ \Phi_{UHECRs}$ as  in previous section
   :$ \Phi_{UHECRs} \simeq 3 \cdot 10^{-18}$ eV) is
   $$ {N_{GLAST}} \sim {247}  $$
   The consequence of this expected signal above the horizons is the
   necessary presence of a background Ultra High Cosmic Rays at a
   rate comparable to present AGASA and HIRES records. The very
   natural advantage is the general calibration of this UHECR physics on
   ground with this high quota Showering in Space. The drawback
   is the need of a clear angle discriminator between HorTaus and
   Hias. Because at the distances we are dealing the split angle
    is nearly one degree we may expect that a dozen or more
    gamma events will be enough to estimate the arrival direction within
a needed accuracy (a few tenth of degree).
  In summary the Glast thresholds are described in the included
  figure below.
\begin{figure}[htbp]
  \epsfig{file=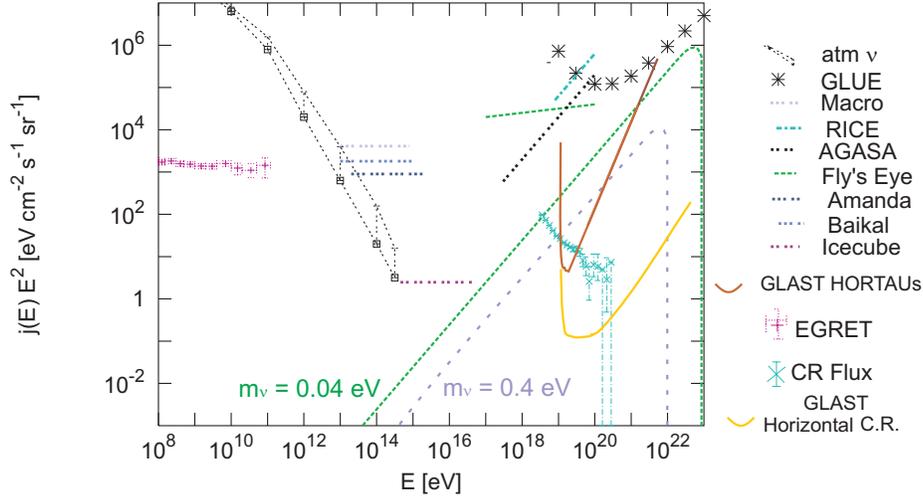,width=1\textwidth,angle=0,clip=}
 \caption{GLAST thresholds for Horizontal Tau air-shower shower,
HORTAUs (or Earth Skimming Showers) over all other $\gamma$, $\nu$
and Cosmic Rays (C.R.) fluence and bounds. The fluence threshold
for Glast has been estimated for a three year experiments
lifetime. Competitive experiment are also shown as well as the
Z-Shower expected spectra in two different light neutrino mass
values ($m_{\nu} = 0.04, 0.4$ eV).  \cite{tau},
\cite{tau2},\cite{Kalashev:2002kx}.}
\end{figure}

\subsection{GLAST}

\begin{figure}
\vspace{0cm}
    \epsfig{file=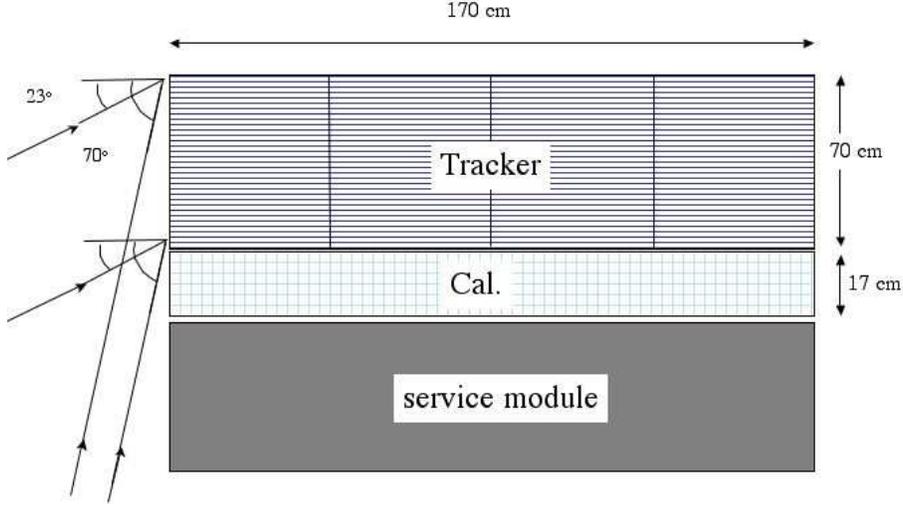,width=1\textwidth,angle=0,clip=}
\caption{\label{glastscheme2} \it Scheme of the lateral view of  GLAST with the arrival directions of horizontal and upward
$\nu\tau$ shower .}
\end{figure}

The Gamma-ray Large Area Space Telescope (GLAST)\cite{glast},  has
been selected by NASA  as a  mission involving an international collaboration of  particle physics and astrophysics
communities from the United States, Italy, Japan, France and Germany  for a launch in the first half of 2006.
The main scientific objects are the study of all gamma ray sources such as blazars, gamma-ray bursts,  supernova
remnants, pulsars, diffuse radiation, and unidentified high-energy sources.
Many years of refinement has led to the configuration of the apparatus shown (see figure~\ref{glastscheme2}),
where one can see the  4x4 array of identical towers each formed by:
$\bullet $   Si-strip Tracker Detectors and converters arranged in 18 XY tracking planes for the measurement
of the photon direction.
$\bullet $ Segmented array of CsI(Tl) crystals for the measurement the photon energy.
$\bullet $ Segmented Anticoincidence  Detector (ACD).
The main characteristics
are an  energy range between    20 MeV and 300 GeV,
a field of view of $\sim$   3  sr,  an energy resolution    of $\sim$ 5\% at 1 GeV, a point source sensitivity of  2x10$^{-9}$ (ph~cm$^{-2}$~s$^{-1}$)     at  0.1 GeV, an event deadtime    of 20 $\mu s$ and a peak effective area  of  10000 cm$^2$,
for a required power    of   600 W and a payload weight of  3000 Kg.

 The list of the people and  the Institution involved in the collaboration together with the on-line
status of the project is available at {\sl http://www-glast.stanford.edu}.

The important number for our estimate is the lateral area of the tracker for each of the four sides that is  $A= 60 cm \cdot 170 cm = 1.02 \cdot 10^4 cm^2$.

The projected total area is $4 \cdot  A* \cos(\theta_{1U}) = 1.4 \cdot 10^4 $
where $\theta_{1U} =70^o$ is the angle between the arrival $\tau$ shower and the horizon constrained by the geometry of the servicing modules that do not allowed to see upward showers (see figure~\ref{glastscheme2})

\section{ Conclusion}

The gamma-ray space experiment  GLAST is just in orbit. Its clear detection of Cosmic rays secondaries, mostly single gamma and electron pairs  as well as muons must take place at a high rate (thousands of events a year). Most muons pairs will hit the detector at 400 GeV energies. More rare  bundle of X-$\gamma$ and $0.4 $ TeV $\mu$ as well as UHE (tens GeV) neutrons (with and without gamma-X traces) might  be also observable soon. PeVs-EeVs cosmic rays air-showering at the terrestrial atmosphere edge must occur at daily-weekly rate in GLAST. The first neutron-gamma-electrons and or muon-gamma-electrons at  associated bundles must flash soon opening a new road to UHECR astrophysics. Moreover with a high angular resolution (below $0.5^o$)it might be even possible in a future to reveal first EeV persistent gamma source as well as rarest PeVs-EeVs upgoing tau.
 This signals are to be distinguished from background noises whose single event or whose rare pair structure in different from rarest (tens) bundles of X-gamma-muons or X-Gamma neutron burst at 0.1 millisecond time structure. This upgoing airshowers will be the most exciting   signal of the long waited UHE Neutrino Astronomy. Similar results , but an order of magnitude below, maybe applied to AGILE detector.


\section{Acknowledgments}


\end{document}